\documentclass[aps,preprint,showpacs,superscriptaddress,groupedaddress]{revtex4-1}  
\usepackage{graphicx}  
\usepackage{dcolumn}   
\usepackage{bm}        
\usepackage{amssymb}   
\usepackage{color}
\usepackage{ulem}

\begin{document}

\title{
Mechanisms of initial oxidation of 4H-SiC (0001) and (000$\bar{1}$) surfaces unraveled by  first-principles calculations
}

\author{Yu-ichiro Matsushita}
\affiliation{Department of Applied Physics, The University of Tokyo, Tokyo 113-8656, Japan}
\author{Atsushi Oshiyama}
\affiliation{Department of Applied Physics, The University of Tokyo, Tokyo 113-8656, Japan}

\date{\today}

\begin{abstract}
We have performed electronic state calculations to clarify the initial stage of the oxidation of the Si- and C-faces in 4H-SiC based on the density-functional theory. We investigate how each Si and C atomic site is oxidized on C- and Si-face, and explore most probable reaction pathways, corresponding energy barriers, and possible defects generated during the oxidation. We have found that carbon annihilation process is different between on Si- and on C-face, and this difference causes different defects in interface; In C-face case, (1), carbon atoms are dissociated directly from the substrate as CO molecules. (2), after CO dissociation, 3-fold coordinated oxygen atoms (called Y-lid) are observed at the interface. (3), high density of C-dangling bonds can remain at the interface. In Si-face case, (1), C atoms inevitably form carbon nano clusters (composed of a few atoms) in interface to reduce the number of dangling bonds there. Moreover, we have found that the carbon nano clusters are composed of not only single but also double chemical bonds. (2), We have observed that CO molecules are dissociated from the carbon nano clusters in MD simulations. Furthermore, we have investigated whether H$_2$ and NO molecules react with the defects found in this study. 
\end{abstract}

\maketitle

\section{Introduction}

Silicon carbide (SiC) is a well known material used for a variety of applications for a long time and today regarded as a promising material for power electronics to sustain our society due to its attractive properties such as wide band gap and high durability under harsh environment \cite{Kimoto_book}.  
One of its advantages over other wide bandgap semiconductors is that insulating films, which are indispensable for electronic devices, are formed by thermal oxidation of SiC, leading to the insulating SiO$_2$ over-layers. This also ensures the good connectivity with the Si technology which would collapse were it not for high-quality ultra-thin SiO$_2$ layers. However, the mobility of current SiC MOSFET (Metal-Oxide-Semiconductor Field-Effect Transistor) is lower  than 10\% of the bulk mobility, indicative of the existence of carrier-trapping defects at the interface\cite{Tilak, Chow, Noborio}. Many experimental and theoretical efforts \cite{Yano, Senzaki, Endo, Dimitrijev, Chung, Kimoto2, Kita, Hijikata, Dhar,Noborito2, Kimoto, Knaup1, Knaup2, Deak, Pantelides,Gavrikov,Li,Pantelides3, Akiyama, Shiraishi} in the past tried to identify such defects at the interface, but the microscopic identification has not been achieved yet. 

Scientifically, oxidation process itself is mysterious: Oxygen attacks SiC to form SiO$_2$, inevitably annihilating a massive amount of carbon atoms \cite{Deal-Grove,kobayashi}. This makes the oxidation of SiC much complicated compared with the Si oxidation. The mechanism of the carbon annihilation is totally unknown, although existence of remaining carbon at the interface as a form of the nano-cluster is speculated in 
some theoretical works \cite{Knaup1, Knaup2, Deak, Pantelides}. 
Another interesting fact related to the thermal oxidation is that the oxidation rate of the C face [i.e., the (000$\bar{1}$) surface] is about ten times faster than that of the Si face [the (0001) surface]. This infers that oxidation processes on the Si- and C-faces are different. Passivation of the carrier-trapping defects also depends on the surface orientation. Hydrogen treatment reduces the defect density in the case of oxidation of C-face. On the other hand, in the case of Si-face, NO, NO$_2$ or POCl$_3$ annealing is used instead of hydrogen treatment\cite{Kimoto2, Noborito2, Okamoto, Pantelides1}. Knowledge of the defect passivation depending on the surface orientation is totally lacking. It is thus highly demanded to clarify the atom-scale mechanism of the SiC oxidation which is associated with the C annihilation, depends on the surface orientation and causes the carrier-trapping defects at the interface.

In this study, we clarify atom-scale processes at the initial stage of the oxidation of the Si- and C-faces in 4H-SiC, based on the density functional theory (DFT)\cite{hohenberg,kohn}. We carry out both static total-energy electronic-structure calculations and molecular dynamics (MD) simulations. We consider how each Si and C atomic site is oxidized and explore most probable reaction pathways and corresponding energy barriers. 

In section \ref{method}, we explain pertinent features of our calculations.  Sections \ref{results_C} and \ref{results_Si} show our results on the C-face, and the Si-face, respectively. We consider the effects of hydrogen and nitrogen treatment after oxidation in section \ref{passiv}. Lastly, we conclude this study in section \ref{sum}.

\section{Calculation conditions}\label{method} 

In our density-functional calculations, we have adopted the Perdew-Burke-Ernzerhof (PBE) functional\cite{PBE} in Generalized Gradient Approximation (GGA) to the exchange-correlation energy functional. It is well known that GGA underestimates band gaps. Hence, when quantitative discussion on band structures is necessary, we have also used a hybrid functional of Hyde-Scuseria-Ernzerhof (HSE06)\cite{Hyde, Kresse, matsushita} to reproduce the values of band gaps correctly.

For the static electronic structure calculations, we have used the Vienna $ab$ $initio$ simulation package (vasp)\cite{vasp}. We have adopted a plane-wave-basis set with energy cutoff of 400 eV to expand the wave function and density. The projector augmented wave (PAW) pseudopotentials \cite{PAW} have been used in this study. In this study we have treated the SiC substrate by a slab consisting of 6 bilayers with the $3 \times 3$ periodicity in the lateral plane. A vacuum region of 10-\AA\ thickness is enough to avoid fictitious interactions between the adjacent slabs. The bottom surface of the slab is fixed to the bulk crystallographic and terminated by H atoms to compensate for the missing bonds, whereas the rest of the system is allowed to evolve and relax freely. Brillouin zone (BZ) integration has been performed with Monkhorst-Pack 2x2x1 sample $k$ points for the slab model. These parameters are adopted after the examination of the accuracy within 0.03 eV in total energy. The structural optimization has been done with a tolerance of $10^{-1}$ eV\AA$^{-1}$. 

In the calculations for molecular dynamics, we have adopted Car-Parrinello MD (CPMD) scheme \cite{CPMD} implemented in real-space DFT code (RSDFT) \cite{RSCPMD,RSCPMD1,RSCPMD2}. In the RSDFT scheme, we introduce real-space grids and all the quantities including Kohn-Sham orbitals and thus valence electron densities are computed on the grid points. The grid spacing has been taken to be 0.19 \AA\ corresponding to the 840-eV cutoff in the plane-wave basis set. All simulations are done within the PBE. BZ integration has been performed with $\Gamma$ point in MD simulations. The ionic temperature is controlled by Nose'-Hoover thermostat \cite{Nose} in the constant NVT simulations.

First, we have explored the most stable surface structure among various stacking sequences and found that the most stable surface stacking sequence is ABC, or $h$ site at the surface. We then consider how each Si and C atomic layer is oxidized on thus determined most stable C- and Si-faces. Namely, we clarify the oxidation process of the topmost surface 3 layers. We have found that the essential features of the oxidation process is common on the surface with different stacking sequence. Stacking sequence generally plays an important role in considering the phenomenon related to conduction electrons because of the floating nature of the conduction electrons \cite{matsushita_floating1, matsushita_floating2}. However, in this study valence electrons are important in oxidation process.

In this study we introduce oxygen molecules near the surface or the interface and then seek for the plausible atomic reactions. To reveal electronic structures, we use the band-unfolding analyses \cite{Zunger2010,Zunger2012,unfolding}. The calculated ``energy bands" for the supercell with the lateral periodicity of 3 $\times$ 3 are unfolded to the BZ of the 1 $\times$ 1 primitive cell. In the unfolding procedure, we calculate the spectral function expressed by the eigen-function of the 1 $\times$ 1 reference system. Hence the obtained unfolding bands in the primitive BZ with their particular spectral weights represent the contribution from the eigenstates of the reference system to the energy spectrum of the target supercell system. 

\section{Oxidation process of C-face}\label{results_C}
In this section, we describe the calculated results of the oxidation process on the C-face. 

\begin{figure}
\includegraphics[width=0.9\linewidth]{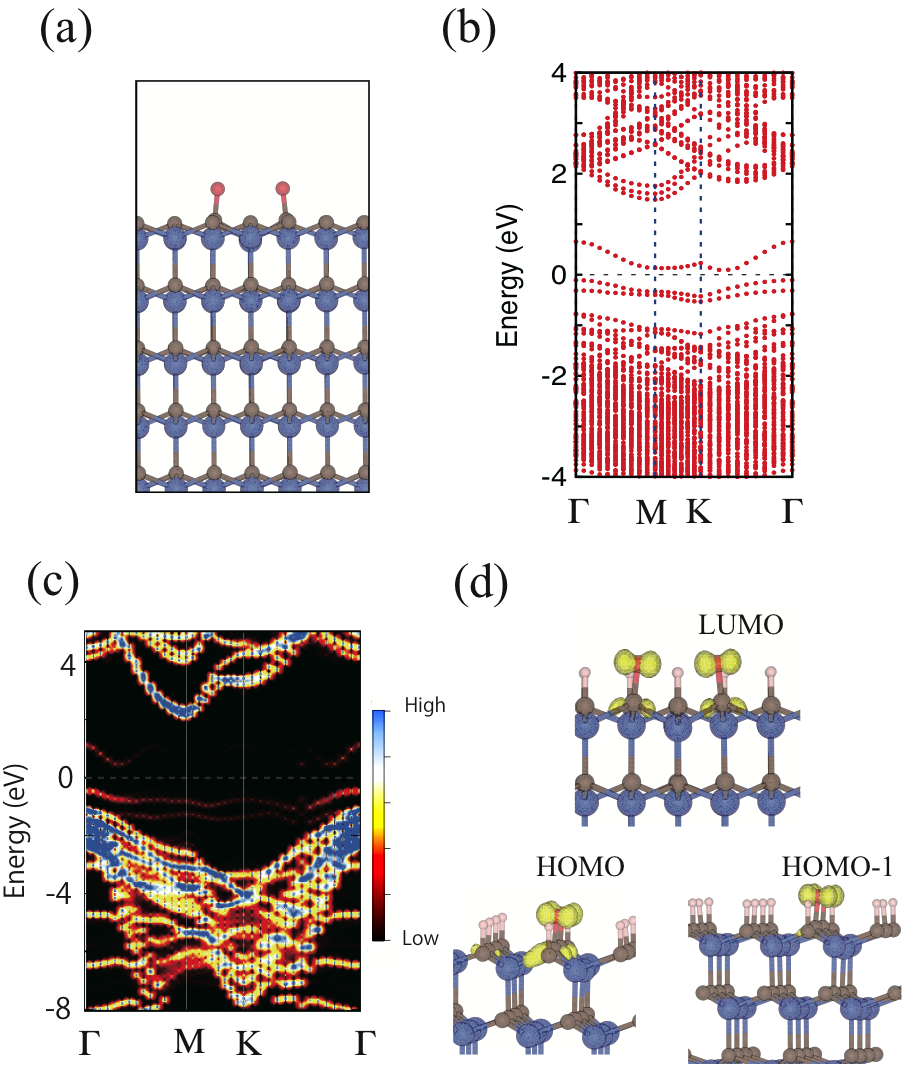}
\caption{(Color online) (a) Optimized structure of two adsorbed oxygen atoms on the C-face. (b) Calculated energy bands of the optimized 3 $\times$ 3 structure with the PBE functional. (c) Energy bands calculated with HSE functional unfolded to the 1 $\times$ 1 BZ. (d), Isovalue surface of the wave functions (Kohn-Sham orbitals) squared at $\Gamma$ point of LUMO (top), HOMO (bottom left), and HOMO-1 (bottom right) with the value of 20 \% of the maximum. The blue, brown, red and white balls depict Si, C, O and H atoms, respectively. The Fermi energy is set to be 0.}
\label{Fig1}
\end{figure}

We start with placing an oxygen molecule on the surface and have optimized the structure. We have found that the molecule is dissociated to two oxygen atoms without energy barrier and that the each of the two O atoms is adsorbed at the on-top position of surface carbon atoms. The most stable configuration of the two oxygen atoms on the surface is the nearest neighbor pair as shown in Fig.~\ref{Fig1} (a). The distance of the two O atoms is 2.65 {\AA}, smaller than that between the two adjacent carbon sites on the surface, i.e., 3.08 \AA. Figure ~\ref{Fig1}(b) shows calculated energy bands with the PBE functional. Here we terminate the surface dangling bonds by hydrogen atoms in order to see clearly the electron states originated from the O atoms. The valence-band maximum (VBM) of 4H-SiC bulk is located at $\Gamma$ point, while the HOMO conduction-band minimum (CBM) is at $M$ point. The calculated band gap of 4H-SiC bulk is 2.3 eV in PBE, which is underestimated because of the well-known drawback inherent in the GGA. It is clear that absorbed oxygen atoms generate three energy levels in the band gap region of 4H-SiC. Figure \ref{Fig1}(d) shows the Kohn-Sham orbitals of those three states, the highest-occupied Kohn-Sham molecular orbital (HOMO), the 2nd HOMO (HOMO-1), and the lowest-unoccupied Kohn-Sham molecular orbital (LUMO).  By the analyses using the tight binding model, HOMO-1 and HOMO are the bonding and the anti-bonding orbitals, respectively, between the oxygen $p$-orbitals extending perpendicular to the O-O direction, and LUMO is the anti-bonding orbital between the $p$-orbitals along the O-O direction. The $p$-orbitals of the oxygen atoms extended along the surface normal direction are strongly hybridized with the top-surface carbon atoms, constituting the bonding and the antibonding states which are located in energy in the valence and conduction bands, respectively. In order to confirm salient features of the energy bands, we have also calculated the band structure with HSE functional and unfolded it to the primitive 1 $\times$ 1 BZ [Fig.~\ref{Fig1}(c)]. The calculated band gap of 4H-SiC is 3.1 eV, which is in quantitatively good agreement with the experimental one, 3.23 eV, and the oxygen-related levels are located in the band gap region $\epsilon ({\rm HOMO}) = \epsilon_{{\rm valence\ top}} + 0.2 eV \pm 0.2 eV$, and $\epsilon ({\rm LUMO}) = \epsilon_{{\rm valence\ top}} + 1.8 eV \pm 0.3 eV$.

\begin{figure}
\includegraphics[width=1.0\linewidth]{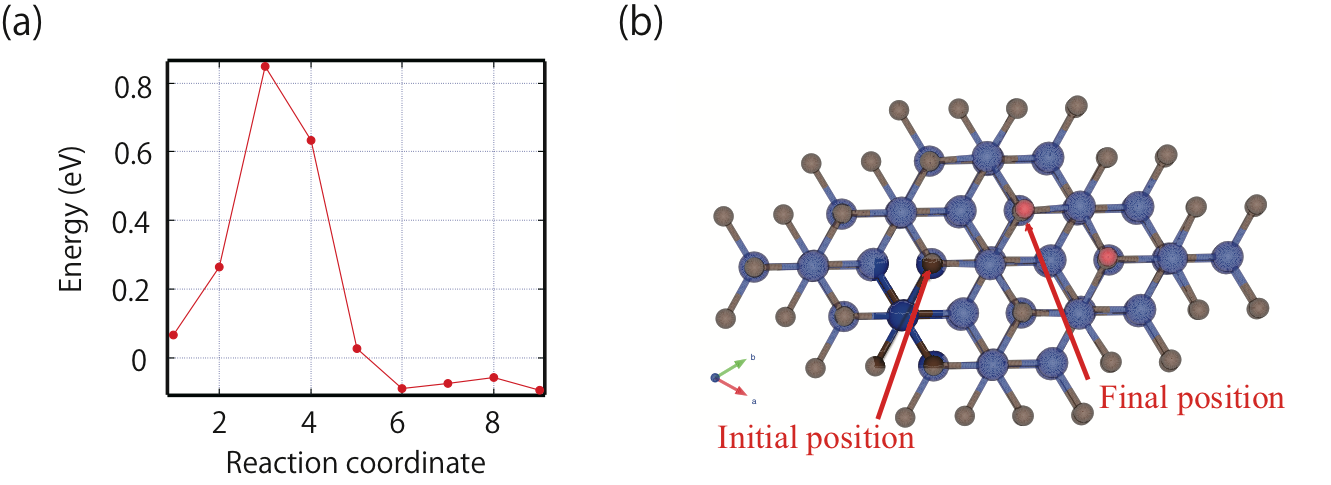}
\caption{(Color online).
(a) Calculated migration barrier of an oxygen atom from the 2nd nearest neighbor to the nearest neighbor site of the other O atom as indicated in (b).}
\label{Fig2}
\end{figure}

The finding that the two oxygen atoms are adsorbed at the nearest neighbor on-top C sites is corroborated by the following additional calculations on the oxygen migration. Suppose the two oxygen atoms adsorbed at the on-top C sites. We consider the energy barrier for one of the two O atoms which is located at the second nearest neighbor of the other to migrate to the nearest neighbor on-top site [Fig.~\ref{Fig2}(b)]. Fig.~\ref{Fig2}(a) shows the obtained energy barrier along the reaction coordinate which is determined by the nudged elastic band method \cite{NEB}. The calculated energy barrier turns out to be 0.78 eV. Also Fig.~\ref{Fig2} clearly tells us that the total energy of the final structure, i.e., a pair of the O atoms located at the nearest-neighbor on-top sites, is lower in energy than the initial structure by 0.1 eV.

\begin{figure}
\includegraphics[width=0.8\linewidth]{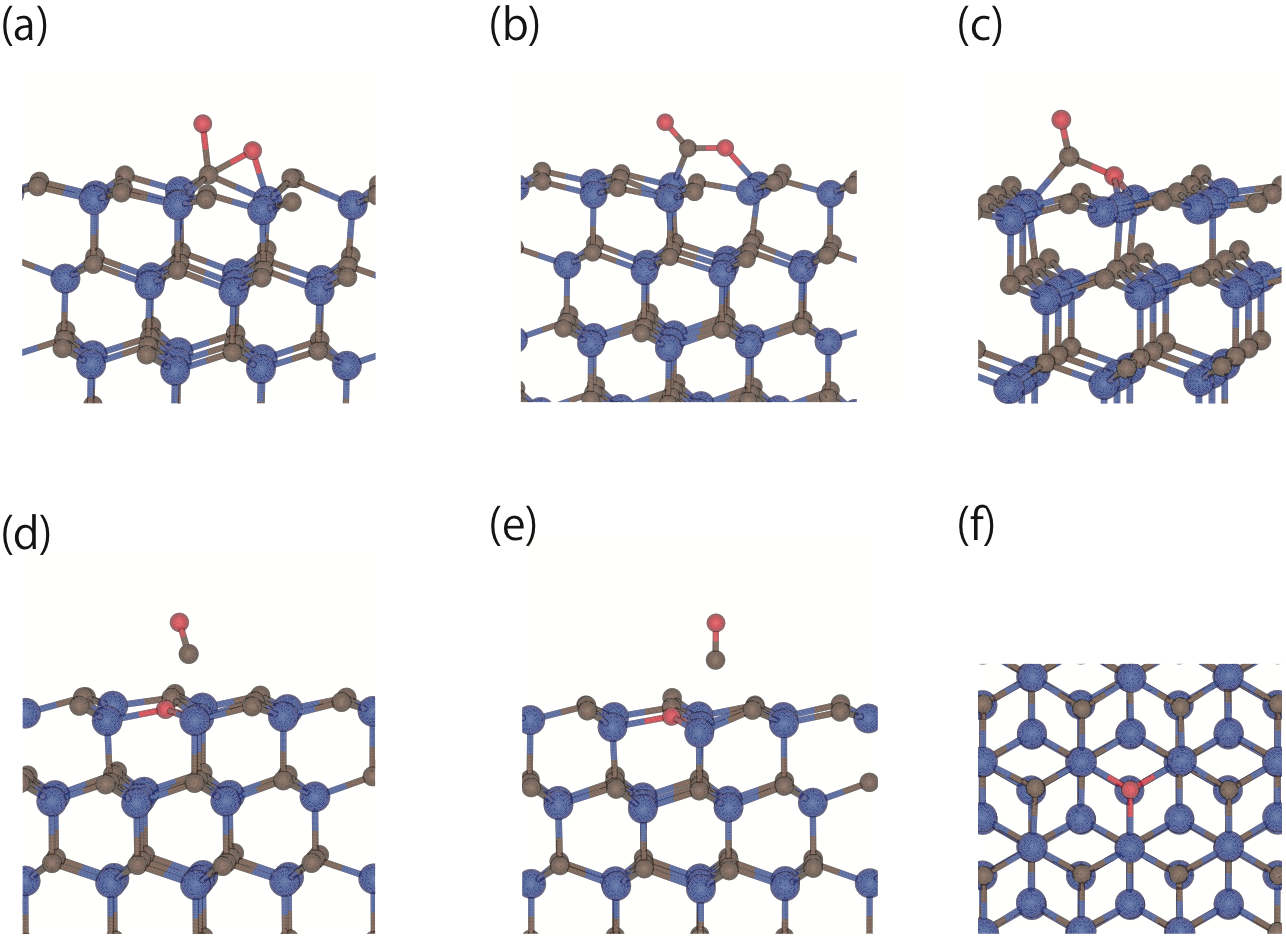}
\caption{(Color online). 
A series of structural optimization after putting two oxygen atoms on a certain carbon site. The initial structure (a) evolves through the structures (b), (c) and (d) and toward the final structure (e). Top view of the SiC surface in the final structure is shown in (f).
}
\label{Fig3}
\end{figure}

The reaction of the O$_2$ dissociation and the subsequent O adsorption on the nearest-neighbor on-top C sites clarified above is expected to occur in a ubiquitous way during the oxidation of the C-faces. Then, we consider a plausible situation in which two oxygen atoms come to a single C on-top site as in Fig.~\ref{Fig3}(a) and perform structural optimization. Then, surprisingly, we have observed that a CO molecule dissociates from the surface as shown in Figs.~\ref{Fig3} (d) and (e). Looking closer at the obtained reaction pathway, the two oxygen atoms and the carbon atom form an isosceles triangle as shown in Fig.~\ref{Fig3}(b). Then, one oxygen atom of the two sinks to the subsurface [Fig.~\ref{Fig3} (c)], making chemical bonds with two silicon and one carbon atoms, while the other oxygen goes upwards followed by the carbon atom, forming a CO molecule [Fig.~\ref{Fig3}(d)]. We have found that a carbon atom on surface is annihilated as a CO molecule, while the other oxygen atom occupies the original carbon site leading to a three-fold coordinated oxygen [Fig.~\ref{Fig3} (f)]. This three-fold-coordinated oxygen occasionally appears in crystalline Si as a plausible form of the oxygen interstitial and called $y$-lid \cite{saito}. We have found that this atomic reaction shown in Figs.~\ref{Fig3} from (a) to (e) gains the energy of 3.2 eV. In contrast to our calculations, the theoretical study in Ref.~\onlinecite{Ono} argues that carbon annihilation takes place as CO$_2$ molecules from their energetic viewpoint. However, we argue here that the oxidation reaction naturally leads to the CO desorption.

\begin{figure}
\includegraphics[width=0.9\linewidth]{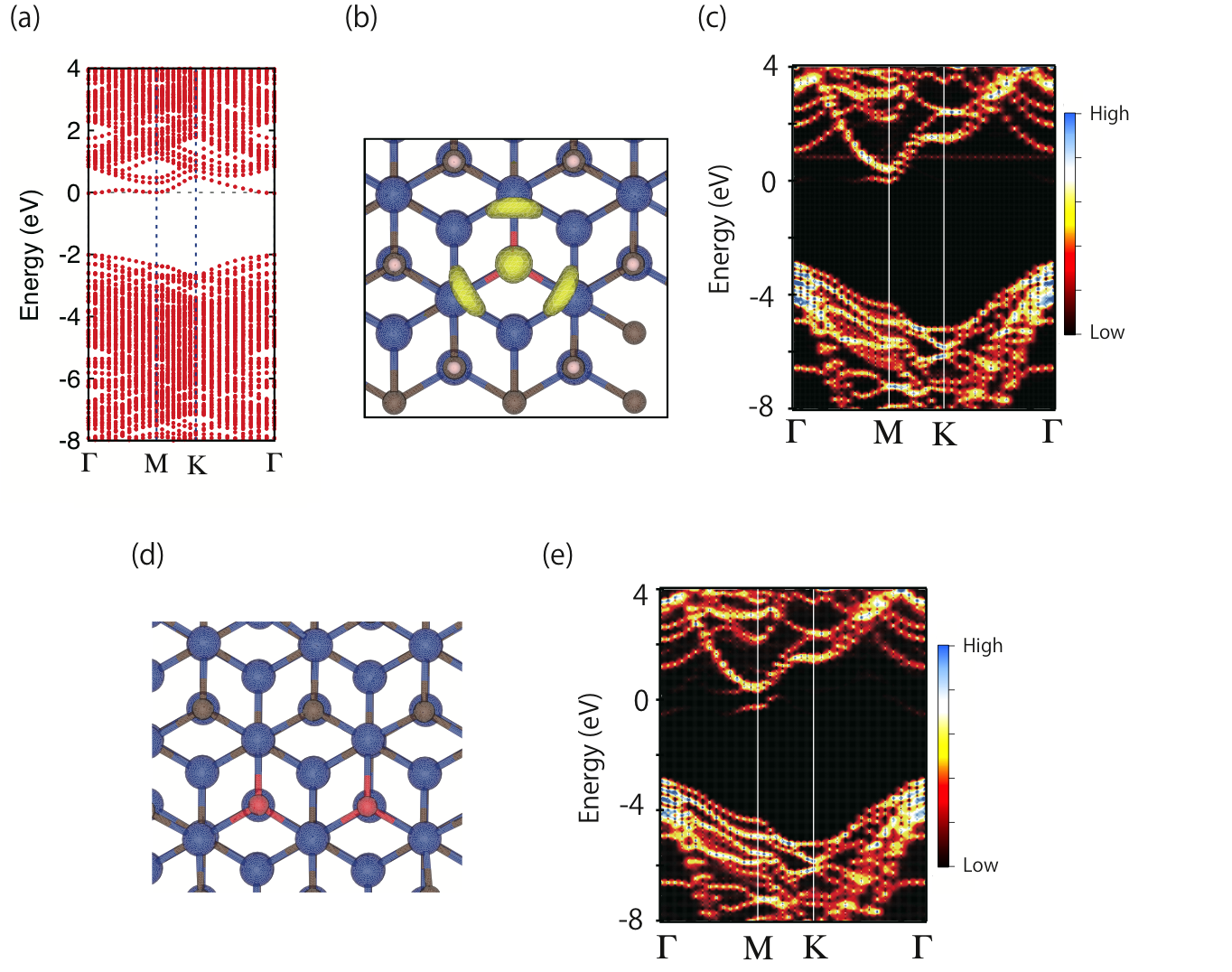}
\caption{(Color online).
Electronic structure of a single $y$-lid found on the SiC surface: Energy bands with PBE functional (a), the isovalue surface of the Kohn-Sham orbital of the gap state at $\Gamma$ (b), and unfolded energy bands with HSE functional (c). Atomic structure (d) and unfolded energy bands with HSE functional (e) of the $y$-lid pair.
}
\label{Fig3_band}
\end{figure}

In Fig.~\ref{Fig3_band}, we show the electronic structure of the $y$-lid on the SiC surface. It is clearly seen that the $y$-lid induces a level just below the CBM. We also show the result with HSE functional in Fig.~\ref{Fig3_band} (c). This electron state below the CBM clearly has an anti-bonding character between an $s$ orbital of the oxygen atom and $p$ orbitals of the surrounding three silicon atoms [Fig.~\ref{Fig3_band} (b)]. We have then added another O$_2$ molecule to the system and examined reactions between the O$_2$ molecule and the $y$-lid. However, the O$_2$ molecule is found to be incapable of destroying the $y$-lid structure. Instead, it is found that another $y$-lid is generated [Fig.~\ref{Fig3_band} (d)], emitting a CO molecule in the same pathway as before. This means that the oxidation process leading to the formation of the $y$-lid structure repeats even in the presence of the $y$-lids, continuing to emit CO molecules. The electronic structure of the $y$-lid pair obtained with the HSE functional is shown in Fig.~\ref{Fig3_band} (e). It is clarified that the two $y$-lids interact with each other, rendering the energy level located near CBM shift downwards by 0.2 eV.

\begin{figure}
\includegraphics[width=1.0\linewidth]{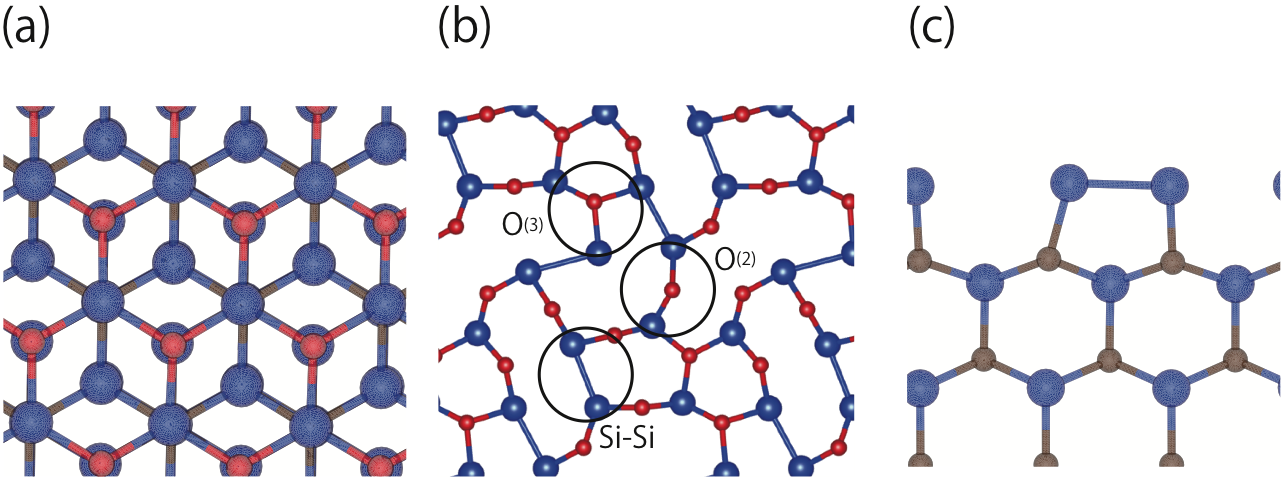}
\caption{(Color online). 
(a) Top view of an initial structure of one-monolayer O-covered surface consisting of $y$-lids. (b) Top view of the two-dimensional amorphous-like network of Si and O obtained by the total-energy minimization of the structure. (c) Side view of the stable structure obtained by the total-energy minimization of the structure (a) after peeling the surface network of Si and C shown in (b).}
\label{Fig4}
\end{figure}

It is therefore highly likely that the surface C atoms are substituted with O atoms one by one during the oxidation. Then, in order to further investigate the oxidation process, we consider a simplified situation where the surface is covered by one-monolayer oxygen with the topmost C being replaced by the O forming the $y$-lid structures as shown in Fig.~\ref{Fig4} (a), and have performed structural optimization. Figures \ref{Fig4} (b) and (c) show the resultant stable geometry in which amorphous-like surface atomic layer of Si and O is formed (Fig.~\ref{Fig4} (b)). Each Si of the layer forms another bond with a sub-surface C atom [Fig.~\ref{Fig4} (c)]. The reason of the formation of an amorphous-like layer is that the Si-O bond is shorter than the Si-C bond and thus the one-monolayer coverage of O plus monolayer Si is insufficient to cover the whole surface. The stress inherent in the structure Fig.~\ref{Fig4} (a) inevitably causes the 2-dimensional amorphous layer of Si and O.

Figure \ref{Fig4} (b) shows a variety of chemical bonds in the surface layer of Si and O. We observe two-fold coordinated oxygen atom (Si-O-Si), denoted by O$_{(2)}$ in Fig.~\ref{Fig4}(b), Si-Si bonds (denoted by Si-Si in the figure), and three-fold coordinated oxygens ($y$-lids) (denoted by O$_{(3)}$ in the figure). Two-fold coordinated oxygen atoms are energetically favorable than the three-fold-coordinated ones. Therefore, some oxygen atoms change their chemical bonds from the three-fold coordination to the two-fold one. However, by changing the chemical bonds around the oxygen atoms (from the three-fold to the two-fold coordinated), some silicon atoms inevitably have dangling bonds. In order to reduce the number of dangling bonds, structural reconstruction occurs on the surface to form the Si-Si bonds. We have found that a silicon atom with dangling bonds goes closer to a nearest neighbor silicon atom, and forms a Si-Si bond. The Si-Si bond length is 2.4 \AA, which is close to the bond length in crystalline Si, i.e., 2.35\AA. As a result, all the Si atoms are free from the dangling bonds in the optimized structure shown in Fig.~\ref{Fig4}(b). They have three chemical bonds in the lateral plane and a single bond with a subsurface C atom. This structural reconstruction forming the Si-Si bonds is reasonable because this decreases efficiently the number of dangling bonds with small distortion energy derived from the slanted Si-C bonds along the z-axis by only 10$^\circ$. 

\begin{figure}
\includegraphics[width=1.0\linewidth]{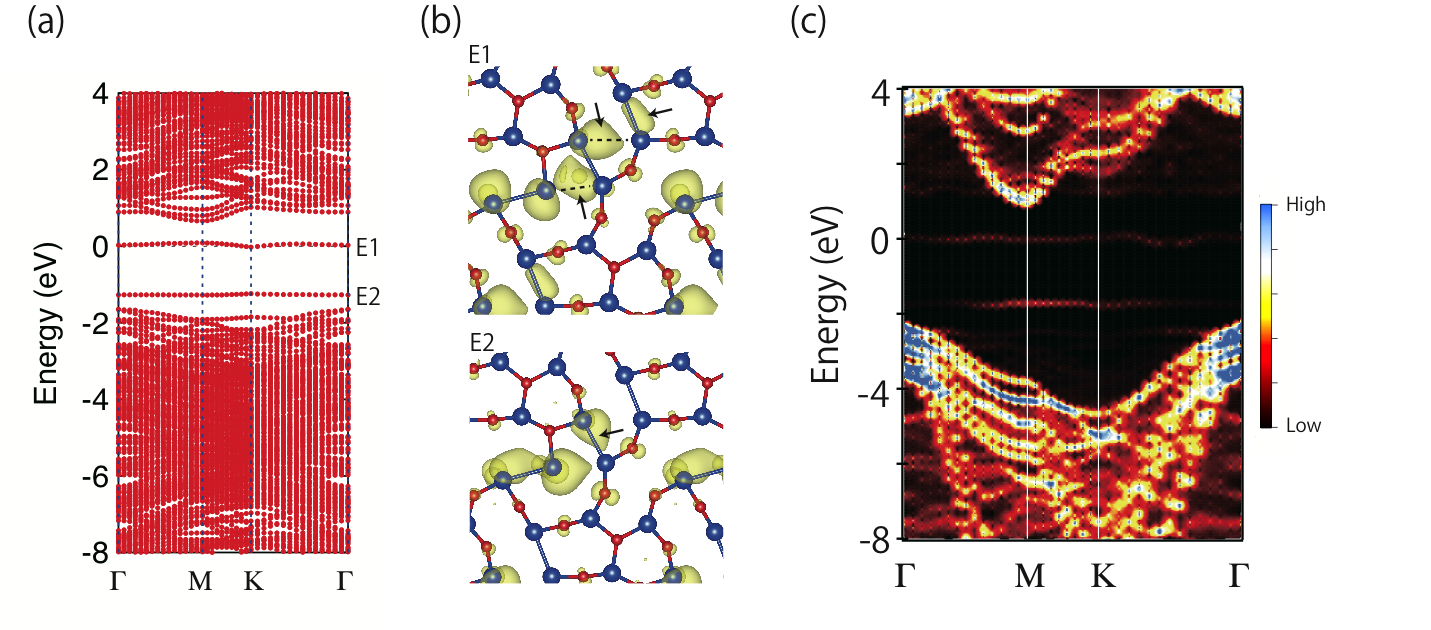}
\caption{(Color online). 
(a) Energy bands of the one-monolayer oxygen-covered amorphous-like structure shown in Figs~\ref{Fig4} (b) and (c) calculated by PBE. (b) Isovalue surfaces of the Kohn-Sham orbitals squared of the gap states E1 (upper panel) and E2 (lower panel). The value adopted is the 10 \% of the corresponding maximum values. Arrows depict the Si-Si bonds along which the Kohn-Sham orbitals are extended. (c) The energy bands calculated with HSE and unfolded to the 1 $\times$ 1 BZ.}
\label{Fig4_band}
\end{figure}

In Fig.~\ref{Fig4_band}, we show the band structure of the above obtained one-monolayer oxygen-covered amorphous-like structure. We have found two electron states in the gap. As shown in Fig.~\ref{Fig4_band} (b), the Kohn-Sham orbitals of the gap states are mainly distributed between the two Si atoms forming the Si-Si bond. The unfolded band structure with HSE is also shown in Fig.~\ref{Fig4_band} (c). The calculated gap states from the Si-Si bonds are located in energy at 0.7eV below the CBM and 0.2 eV above the VBM.

\begin{figure}
\includegraphics[width=0.7\linewidth]{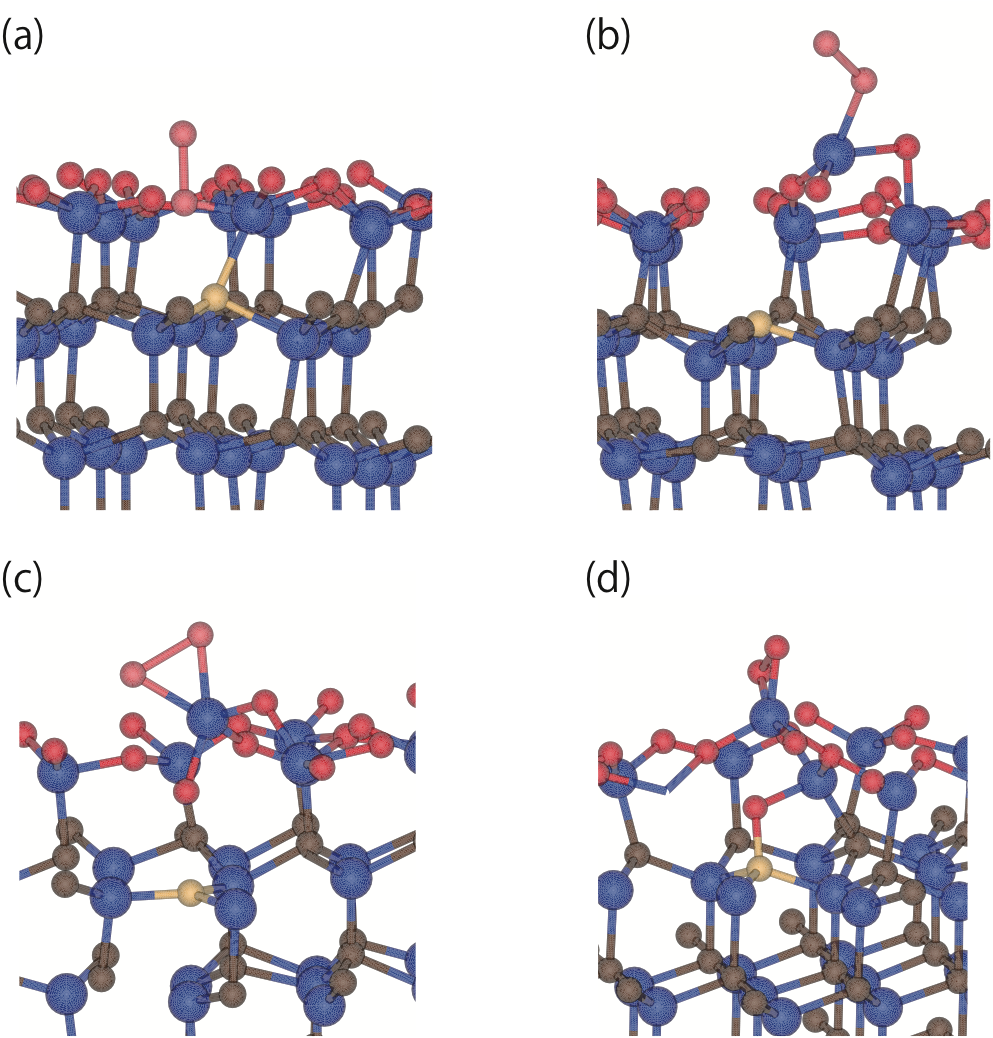}
\caption{(Color online).
Time evolution [from (a) to (c)] in the CPMD simulation at 1300 K after adding an O$_2$ molecule near a Si atom bonded with 3 O atoms in the one-monolayer-oxygen-covered structure shown in Fig.~\ref{Fig4} (b). (a) Initial structure in which the Si atom is surrounded by 4 O atoms. The O$_2$ molecule added are depicted by pink balls. In (b) and (c), the Si atom makes four Si-O bonds, causing a dangling bond at the subsurface C atom depicted by the yellow ball.
(d) The carbon atom with a dangling bond catches an additional O atom making a C-O-Si bond.}
\label{Fig5}
\end{figure}

\begin{figure}
\includegraphics[width=0.9\linewidth]{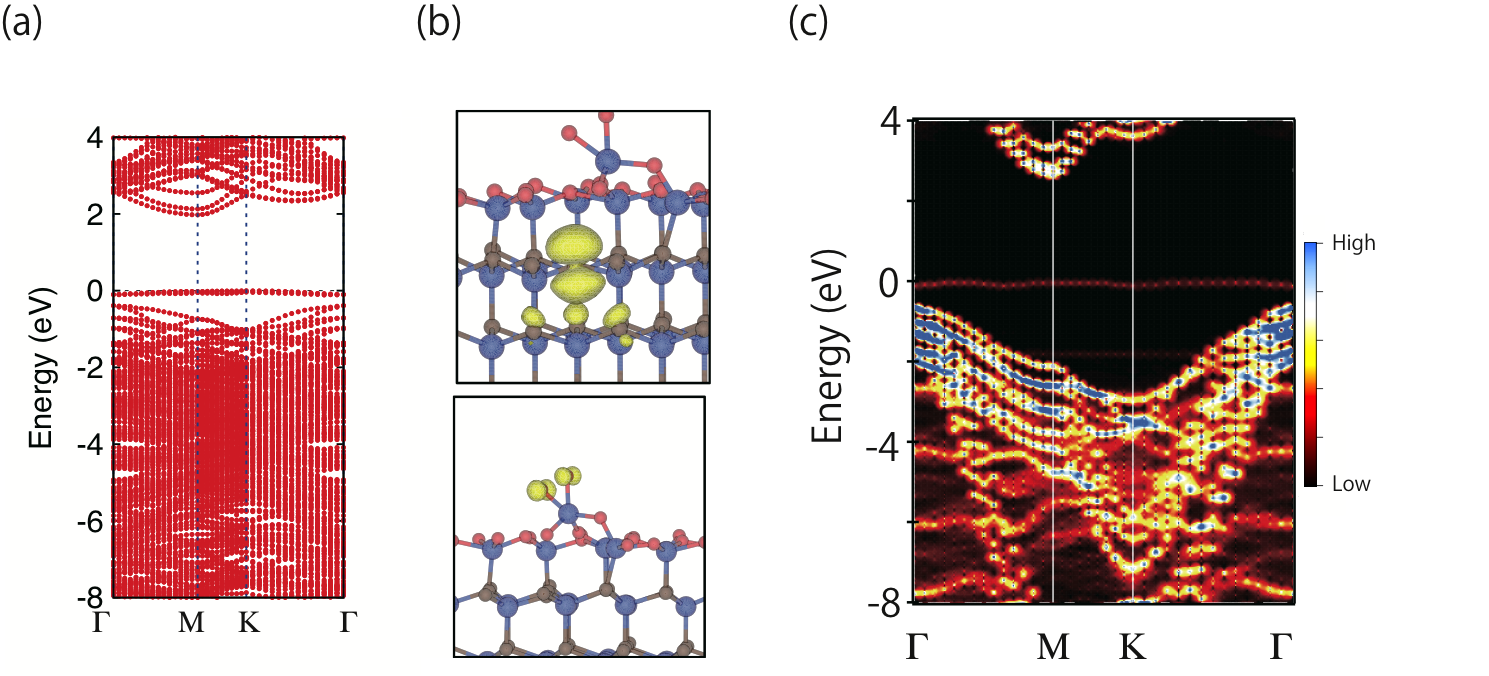}
\caption{(Color online). 
Energy bands of the structure shown in Fig.~\ref{Fig5} (c) calculated by the PBE functional (a) and by the HSE functional unfolded to the primitive BZ (c). The Kohn-Sham orbitals of the two gap states almost degenerate above VBM. The orbitals squared are shown as their isosurfaces with the values of 50 \% and 20 \% for the upper and lower panels, respectively, of their maximum values.
}
\label{Fig5_band}
\end{figure}

We next introduce an O$_2$ molecule in this one-monolayer-oxygen-covered surface and optimize the structure. We have found that when the oxygen molecule is near the Si-Si bond, it is dissociated and each O atom forms a Si-O bond. On the other hand, when the added oxygen molecule attacks an Si atom which is bonded with three O and a single C atoms [Fig.~\ref{Fig5}(a)], the bond between the Si atom and the subsurface C atom is broken and instead 4th Si-O bond with the O atom of the added molecule is formed [Fig.~\ref{Fig5}(b)]. This bond destruction and formation is consistent with our calculations that the bonding energy between Si and O is 4.60 eV, being larger than that between Si and C of 3.49 eV. The bond length between Si and O, 1.7 \AA, is shorter than the length between Si and C, 1.9 \AA. This mismatch of the bond length by as much as 10\% inevitably causes large distortion near the interface. The stress induced by this distortion makes the silicon bonded with 4 oxygen atoms ejected out from the surface, leaving a dangling bond at the subsurface C atom (the yellow ball in Fig.~\ref{Fig5}(c)). Figures \ref{Fig5_band} (a) and (c) show the energy bands calculated with the PBE and the HSE functionals, respectively. Two gap states appear at almost the same energy, i.e., 0.2 eV by PBE or 0.3 eV by HSE above the VBM. One state has a character of the dangling bond of the carbon atom, whereas the other state has a character of the lone pair of the oxygen atom, as is clearly shown in the Kohn-Sham orbital in Fig.~\ref{Fig5_band} (b). 

\begin{figure}
\includegraphics[width=0.8\linewidth]{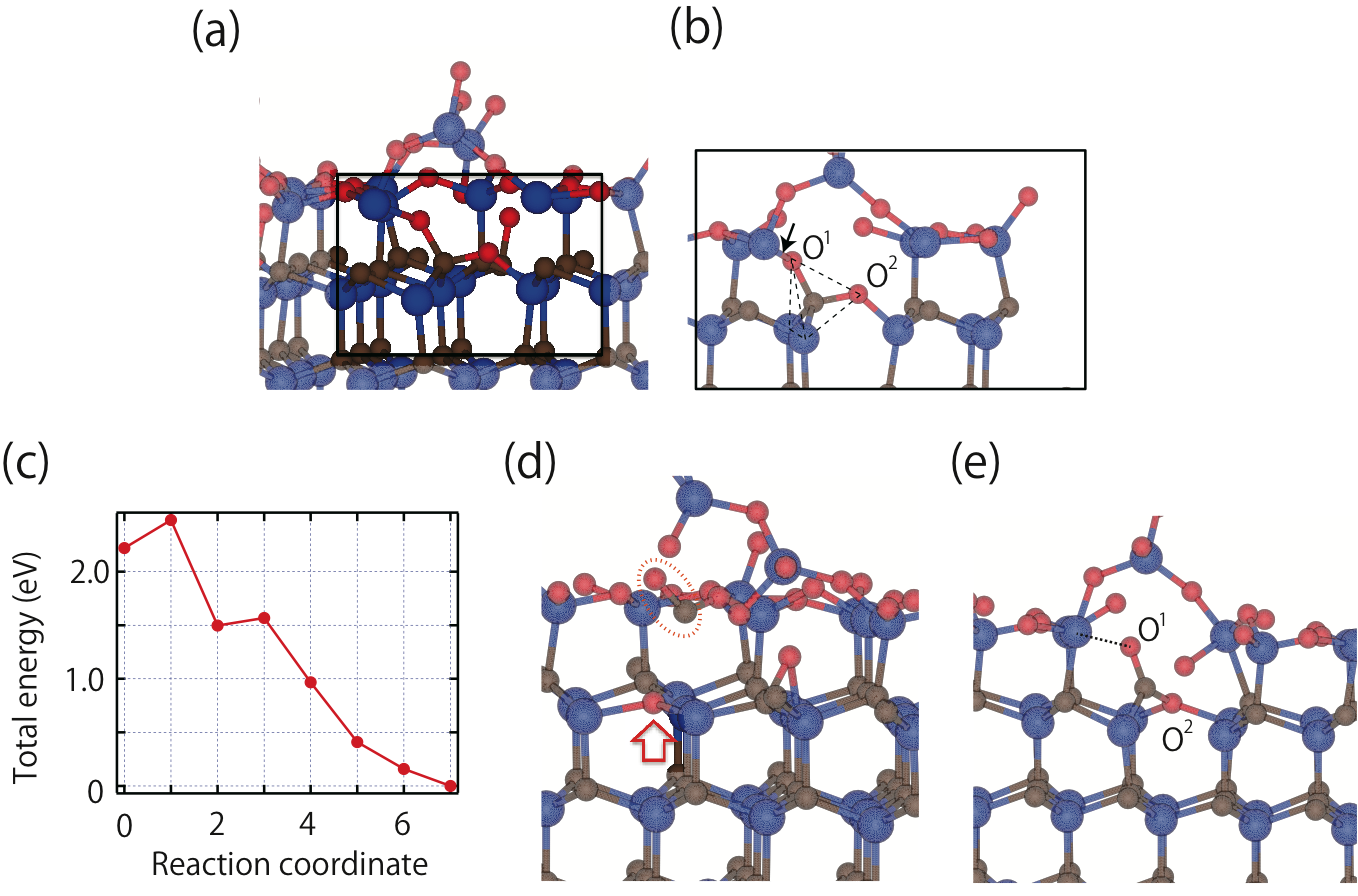}
\caption{(Color online). 
Oxidation process of the 2nd carbon layer on the C-face. Obtained local minimum structure of a carbon atom bonded with two oxygen atoms (a). Squared area is enlarged in (b). (c) The calculated energy barrier of the CO desorption from the local minimum structure shown in (a). (d) The final structure at the reaction coordinate 7 where the $y$-lid and the CO molecule are marked. In (e) the structure at reaction coordinate 2 is shown. The distance of the dotted line is 2.13\AA.
}
\label{Fig6_pre}
\end{figure}

The carbon atom with a dangling bond is expected to be the next reaction center. When a next oxygen molecule approaches the carbon, we have indeed found that the carbon makes a bond with the oxygen atom easily (Fig.~\ref{Fig5}(d)). 
Furthermore, when another oxygen molecule attacks the C-O part (shown by the yellow ball in Fig.~\ref{Fig5}(d)), the oxygen molecule is dissociated, and one O atom is adhered to the carbon atom, and finally two oxygen atoms stick to this carbon atom, being located at on-bridge positions between the C and adjacent Si atoms. The obtained structure shown in Figs.~\ref{Fig6_pre} (a) and (b) is locally similar to the structure before the CO desorption on the topmost layer [Fig.~\ref{Fig3} (a)]. Then we explore reactions for the CO desorption from this subsurface bond network using the Nudged elastic band method. Initial structure is of course the structure shown in Fig.~\ref{Fig6_pre} (a). The final structure is shown in Fig.~\ref{Fig6_pre} (d) in which the CO unit escapes from the subsurface region and is located at the interstitial site near the surface. Fig.~\ref{Fig6_pre} (c) shows the energy profile along thus determined reaction coordinate. In determining the reaction coordinate, we introduce six hyperplanes during the initial and final geometries and use the nudged-elastic scheme. We have found the rate-determining barrier of 0.3 eV at the geometry on the first hyperplane (reaction coordinate 1). In this transition state, an oxygen atom (labeled by O$^1$ in Fig.~\ref{Fig6_pre} (b)) breaks a Si-O bond (pointed by an arrow in the Fig.~\ref{Fig6_pre} (b)), i.e., the distance between Si and O increases from 1.73 {\AA} to 1.89 \AA. At the reaction coordinate 2, the Si-O distance increases to 2.13 \AA, and the other oxygen atom (labeled by O$^2$) forms two chemical bonds with silicon atoms beneath [Fig.~\ref{Fig6_pre} (e)]. After the the desorption of the CO molecule, the remained oxygen atom [labeled by O$^2$ in Fig.~\ref{Fig6_pre} (e) and also pointed in Fig.~\ref{Fig6_pre} (d)] forms a $y$-lid structure. We have found that this is a similar reaction as the oxidation of the topmost C-layer shown in Fig.~\ref{Fig3}. The total energy gain of this process is calculated to be 3.0 eV, comparable with the corresponding energy gain associated with the topmost C-layer oxidation.

We have now clarified the oxidation process of the top-layer C, the next-layer Si and the third-layer C sites. From the characteristics of the obtained reaction pathways and the corresponding energy barriers, we expect that similar oxidation processes are relevant in the oxidation of the subsequent layers. 

\section{Oxidation process of Si-face}\label{results_Si}
\begin{figure}
\includegraphics[width=1.0\linewidth]{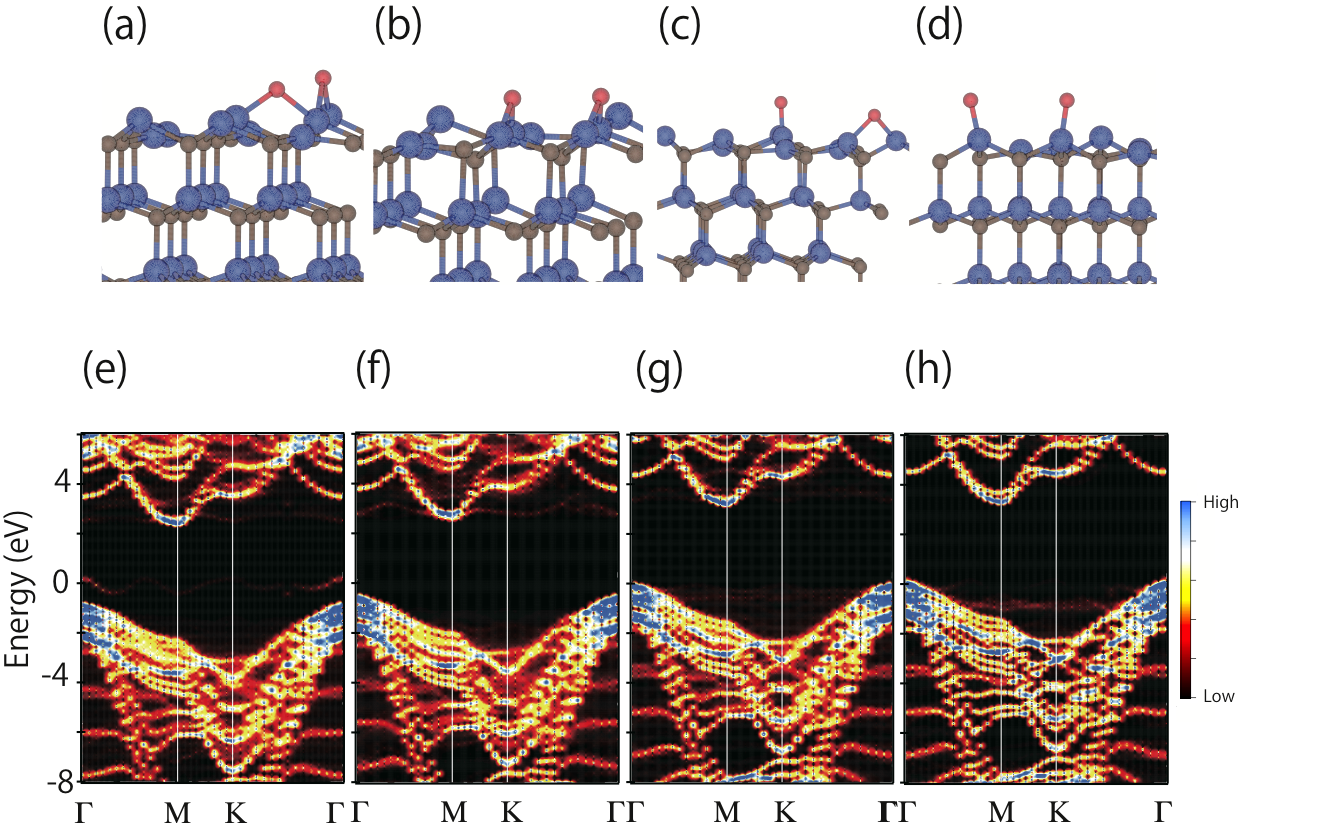}
\caption{(Color online). 
Several oxygen configurations (upper panels), and calculated band structures (lower panels) on Si-face. Each lower panel corresponds to the respective upper panel. }
\label{Fig7}
\end{figure}

In this section, we discuss the oxidation process on Si-face. First we have explored the stable configurations of two oxygen atoms on the surface. We have found that 4 surface morphologies are stable shown in Fig.~\ref{Fig7}. Each oxygen atom can take one of the two different kinds of configuration; one is on-top site on a silicon atom, and the other is on-bridge site between two silicon atoms. We have compared the total energy for the 4 configurations. The most stable structure is the Fig.~\ref{Fig7}(b) in which both two oxygen atoms occupy on-bridge sites. The total energy of the other structure is 0.13 eV higher for Fig.~\ref{Fig7}(a), 0.327 eV for Fig.~\ref{Fig7}(c), and 0.984 eV for Fig.~\ref{Fig7}(d) than that of Fig.~\ref{Fig7}(b). We can see a clear tendency that on-bridge sites are energetically favorable than on-top sites. 

We have also calculated the electronic band structures of them, in which the surface dangling bonds are terminated by H atoms to see clearly the electron state originated from O atoms (The lower panels in Fig.~\ref{Fig7}). We can see strong dependence of the band structure on oxygen positions. We can see a mid-gap state 0.5 eV above the VBM in Fig.~\ref{Fig7}(e). In contrast, in Fig.~\ref{Fig7}(f), (g), and (h) there are no mid-gap states in gap region. 

It is therefore highly likely that the number of on-bridge oxygen atoms increases with further progress of oxidation of the topmost Si-layer. In order to further investigate the oxidation process of the next carbon layer, we adopted a simplified model where the topmost silicon sites are fully oxidized, and covered by one-monolayer oxygen. Then, we put oxygen atoms as many as possible at on-bridge positions of the surface, and we have performed structural optimization. However, the structure doesn't change so much from the initial structure, and each silicon atom in surface forms chemical bonds with single O and three subsurface C atoms.

\begin{figure}
\includegraphics[width=0.7\linewidth]{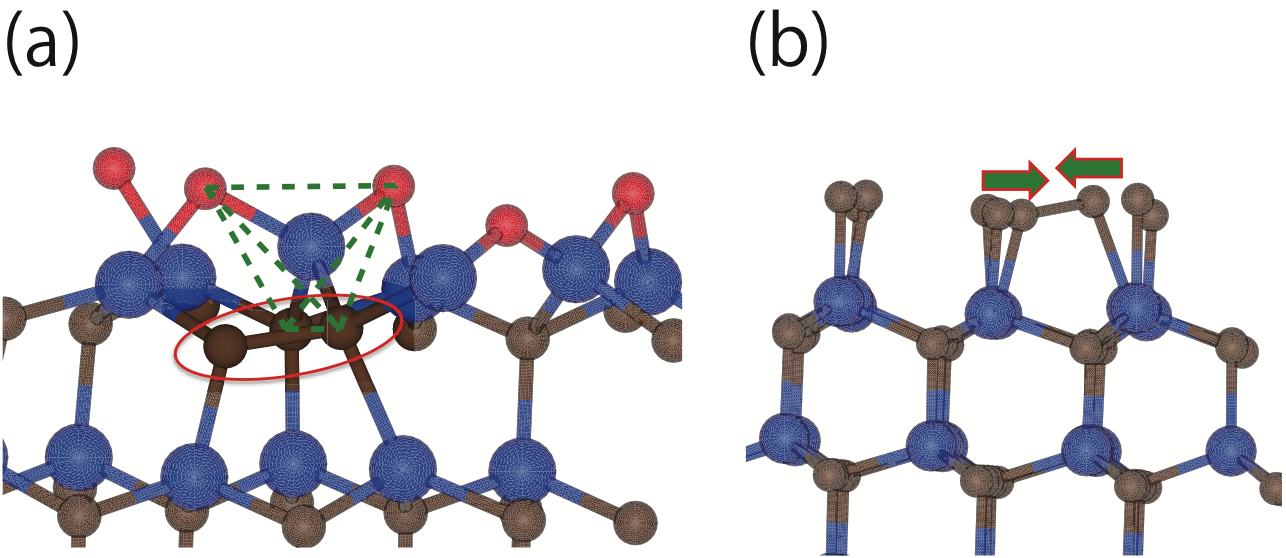}
\caption{(Color online). 
(a), Side view of the obtained structure by total-energy-minimization of the structure after adding an O$_2$ molecule near the one-monolayer-oxygen-covered surface. A silicon atom is surrounded by two oxygen and two carbon atoms (marked by green dotted lines). (b), Side view of the obtained structure shown in (a) after peeling the surface network of Si and O.}
\label{Fig9}
\end{figure}

We further introduced another O$_2$ molecule in this one-monolayer-oxygen-covered surface to proceed the oxidation. By adding an O$_2$ molecule, the bond between a Si and the subsurface C atom is broken, and 2nd Si-O bond with the O atom of the added molecule is formed (Fig.~\ref{Fig9}(a))). As a result, two Si atoms get surrounded by two oxygen atoms. This bond destruction and formation inevitably cause a dangling bond at the subsurface C site. The silicon atom with two Si-O bonds moves slightly upward from the surface. At the same time, the carbon atom with the dangling bond gets closer to a nearest neighbor carbon atom to make a new chemical bond with. Finally, a carbon dimer, C-C bond, appears at the surface, and reduces the number of dangling bonds at C sites (See Fig.~\ref{Fig9}(b)). This reconstruction is similar to C-face case, where Si-Si bonds are formed to reduce the dangling bonds at Si sites (Fig.~\ref{Fig4}(e)). This energy gain due to the formation of C-C bonds indicates that C-C bonds are inevitable during the oxidation on Si-face.

\begin{figure}
\includegraphics[width=0.7\linewidth]{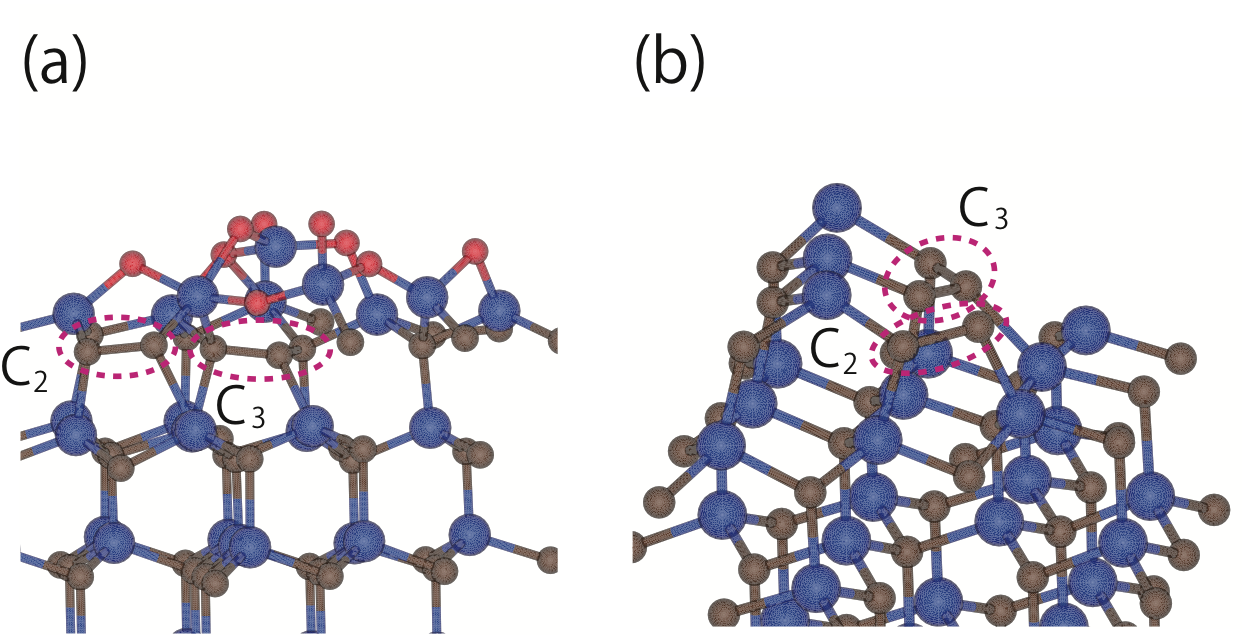}
\caption{(Color online). 
(a) Carbon dimer (C$_2$) and trimer (C$_3$) at the interface generated during the oxidation. (b) The obtained structure shown in (a) after peeling the surface network of Si and O.}
\label{Fig10}
\end{figure}

We proceeded the oxidation further by adding another O$_2$ molecule. Then, it is observed that two surface Si atoms repeat the bond destruction and formation process explained above, forming a carbon dimer (C$_2$) and a trimer (C$_3$) (See Fig.~\ref{Fig10}(a)). Surprisingly, it is noteworthy that the carbon nano clusters (C$_2$ and C$_3$) are composed of not only $sp^3$ but also $sp^2$ bonds. In Fig.~\ref{Fig10}(b), we show the interfacial structure after peeling the surface network of Si and O. We can see clearly $sp^2$ bonds at C$_3$ cluster. 

\begin{figure}
\includegraphics[width=0.9\linewidth]{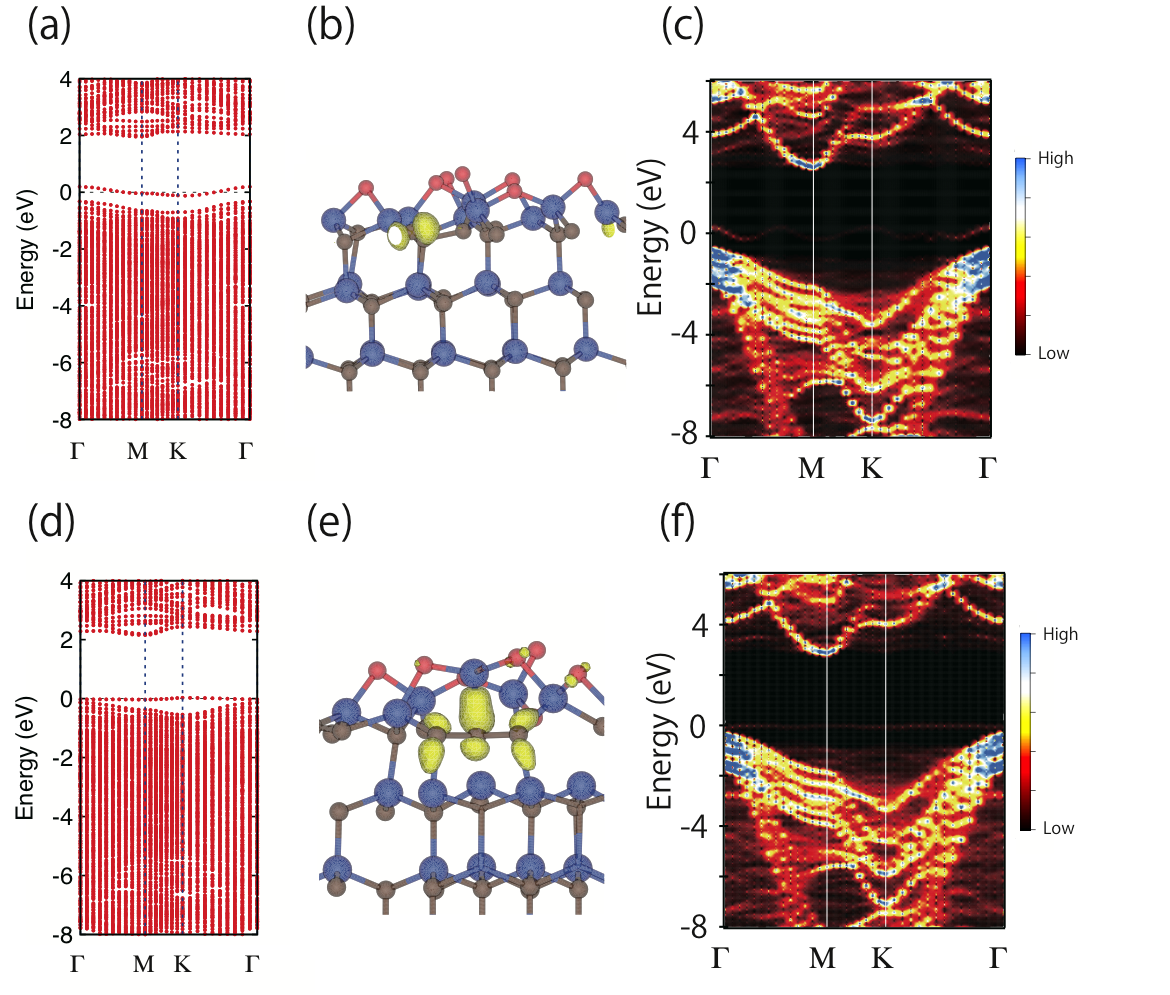}
\caption{(Color online). 
Electronic structure for the structure of Fig.~\ref{Fig9} calculated by PBE (a), and by HSE (c). Electronic structure for the structure of Fig.~\ref{Fig10} calculated by PBE (d), and by HSE (f). 
(b) and (e) Isovalue surface of the wave function of the gap state shown in (a) and (d), respectively. }
\label{band_CC}
\end{figure}

We have also calculated electronic band structures for the carbon nano clusters (Fig.~\ref{band_CC}). The electronic band structure for the structure of Fig.~\ref{Fig9}(a) is shown in Fig.~\ref{band_CC}(a) by the PBE functional, and in Fig.~\ref{band_CC}(c) by the HSE. We can clearly see a mid-gap state 0.6 eV above the VBM by the HSE. Its wave function is shown in Fig.~\ref{band_CC}(b). The wave function mainly contributes to the C$_2$. Fig.~\ref{band_CC}(d) shows the calculated band structure of the structure of Fig.~\ref{Fig10}. A localized state around the C$_3$ (Fig.~\ref{band_CC}(e)) is observed just above the VBM. The character of the wave function is carbon $\pi$ orbitals in the C$_3$ cluster. The unfolded band structure is shown in Fig.~\ref{band_CC}(f). 

\begin{figure}
\includegraphics[width=0.6\linewidth]{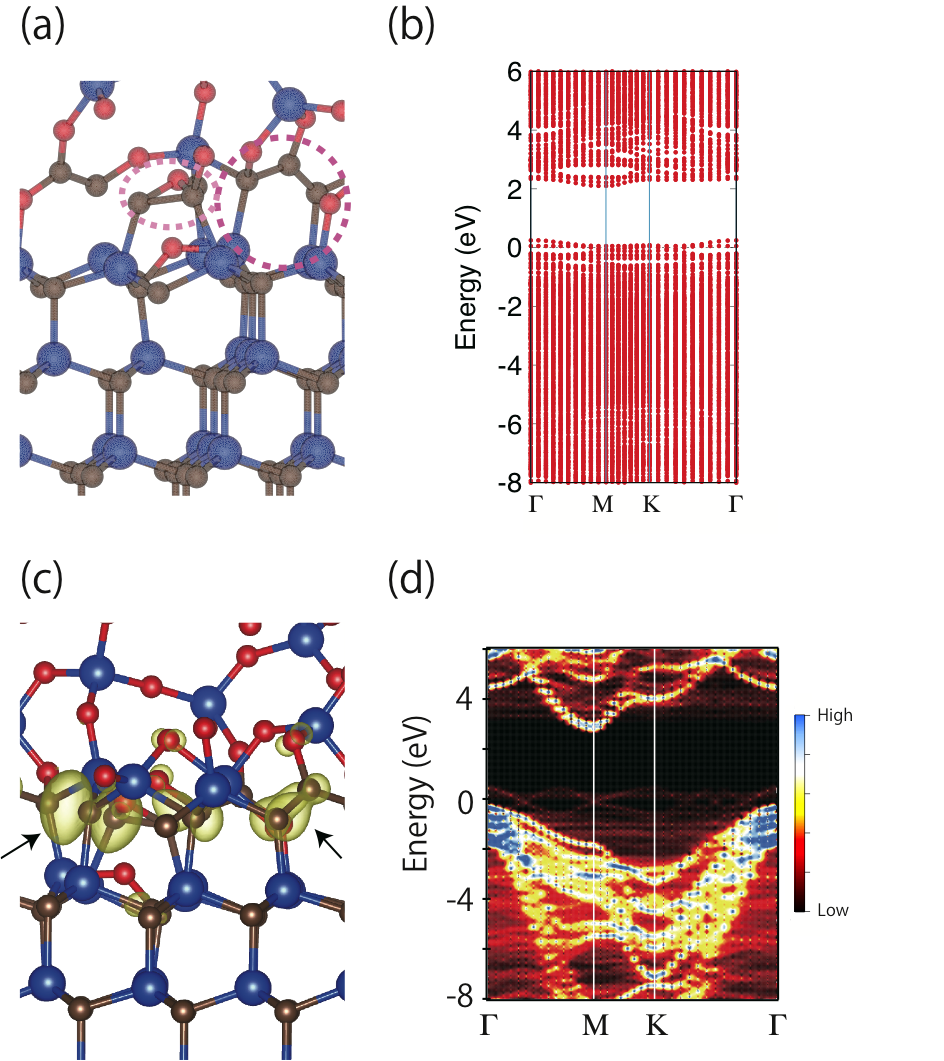}
\caption{(Color online). 
(a) Carbon dimer (C$_2$) and trimer (C$_3$) at the interface generated during the oxidation. (b) Electronic band structure calculated by PBE. (c) Isovalue surface of the wave function at 0.2 eV above the Fermi energy at 5\% of the maximum values. In (d), calculated unfolded band structure with HSE is shown.}
\label{Fig14}
\end{figure}

We repeated adding an O$_2$ molecule and optimizing the structures to proceed the oxidation further. The yielded structure is shown in Fig.~\ref{Fig14}. We can see C$_2$ and C$_3$ at the interface. Another important fact is that we couldn't see any carbon annihilation. This is probably because the C-C bonds of C$_2$ and C$_3$ are stronger than Si-C bonds; i.e., bond strength of C=C (C-C) is 6.07 (3.65) eV, in contrast to Si-C is 3.47 eV. This result indicates the possibility of the existence of C$_2$ and C$_3$ in interface. 

\begin{figure}
\includegraphics[width=0.9\linewidth]{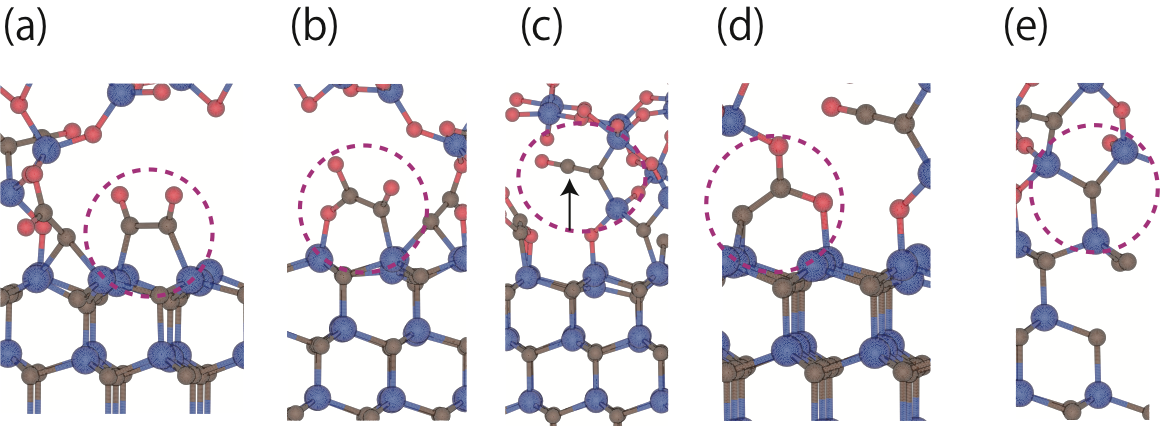}
\caption{(Color online). (a) Frequently observed carbon-related defect structures during the MD simulation at 1600 [K]. In (c), An arrow represents a two-fold coordinated C atom.}
\label{Fig15}
\end{figure}

To investigate the thermal stability of carbon nano clusters, we have performed CPMD simulations at the simulation temperature of 1600 [K]. We couldn't observe the destruction process of the carbon nano clusters within the simulation time of 20 psec. Furthermore, we have found that the size of carbon nano clusters doesn't change also in the MD simulations; each carbon nano clusters are composed of two or three C atoms. In this simulation, we observed that two C$_2$ clusters were combined at one moment. However, the C$_4$ cluster was decomposed into two C$_2$ clusters soon, and thus we couldn't see the process of their growing bigger. This fact might show that larger carbon nano clusters, composed of more than two or three C atoms, are unstable. We took a structure at an instant of the MD simulation, and optimized the structure to obtain the carbon-related-defect interface (See Fig.~\ref{Fig15}). These defects were all frequently observed in the MD simulation. As clearly seen in Fig.~\ref{Fig15}, most C atoms in the interface form C$_2$ nano clusters. All the carbon atoms in the interface are three-fold coordinated except for the two-fold coordinated carbon atom pointed by an arrow in Fig.~\ref{Fig15}(c) (This structure is also reported in \cite{Pantelides2}). We can see ethylene-like structures shown in Fig.~\ref{Fig15}(a) and (b), line-shaped C$_2$O structure in Fig.~\ref{Fig15}(c), $sp^2$ and $sp^3$-combined C$_2$ cluster in Fig.~\ref{Fig15}(d), and mono C atom embedded in SiO$_2$ in Fig~\ref{Fig15}(e). 

\begin{figure}
\includegraphics[width=0.9\linewidth]{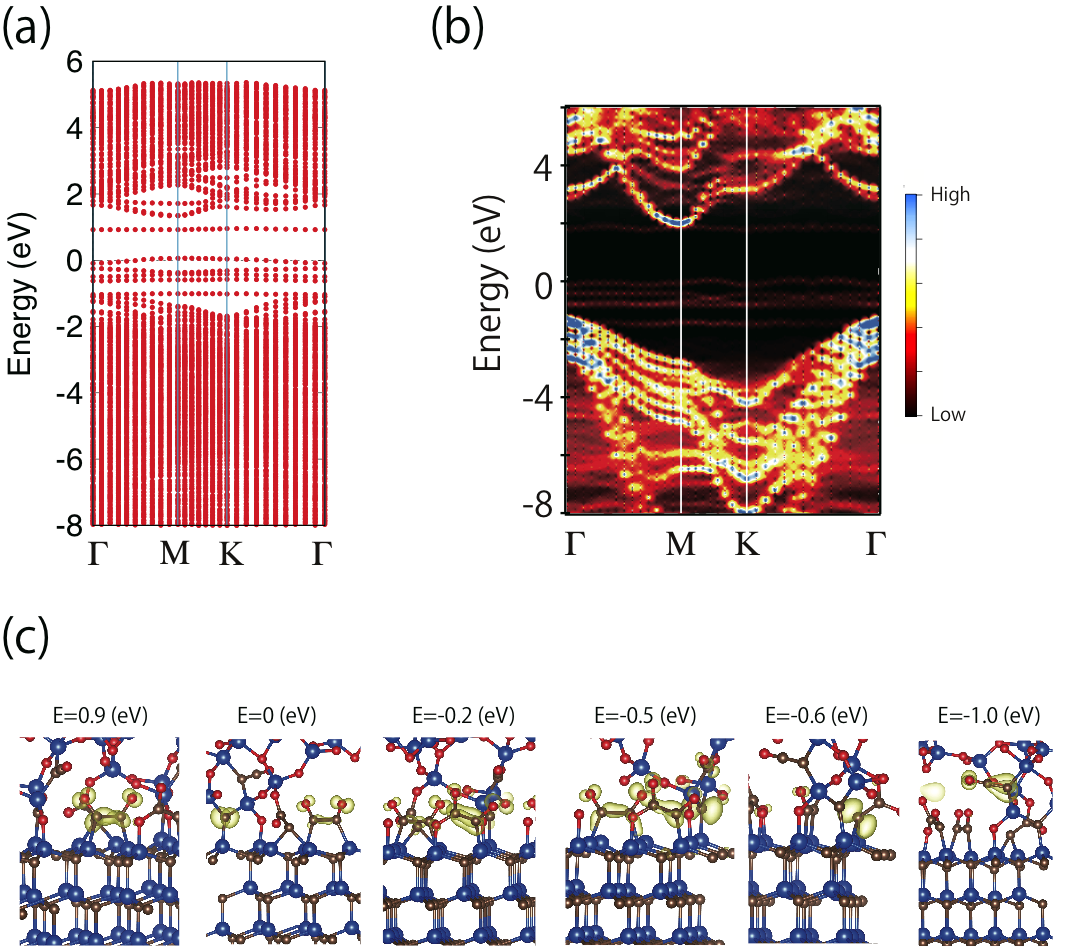}
\caption{(Color online). (a) Calculated band structure with PBE. (b) Electronic band structure calculated by HSE is unfolded to the primitive BZ. (c) Iso-value surface of the wave functions of the gap states at 10\% of maximum values.}
\label{Fig16}
\end{figure}

The calculated band structure of the carbon-related defects are shown in Fig.~\ref{Fig16}. As is clearly seen in Fig.~\ref{Fig16}(a), an electronic state appears near CBM, four states at E=0 $\sim$ $-$0.6 (eV), and one state just above the VBM. The unfolded band structure with the HSE is shown in Fig.~\ref{Fig16}(b). Fig.~\ref{Fig16}(c) shows the wave function of the each defect level. The character of the defect level near the CBM is the $\pi$ state of ethylene-like C$_2$ cluster (Fig.~\ref{Fig15}(a)) exhibiting anti-bonding character between the C and the O atoms. The mid-gap state at E=0 (eV) has an amplitude at the mono carbon site (Fig.~\ref{Fig15}(e)) with $\pi$ character, and at ethylene-like C$_2$ cluster (Fig.~\ref{Fig15}(a)) with $\sigma$ character. The electron states around E= $-0.2 \sim -0.6$ (eV) have characters of $\pi$ and $\sigma$ orbitals of Fig.~\ref{Fig15}(b) and C dangling bond of the structure Fig.~\ref{Fig15}(d). The gap state near the VBM is the $\pi$ orbital of the structure Fig.~\ref{Fig15}(c). 

\begin{figure}
\includegraphics[width=0.9\linewidth]{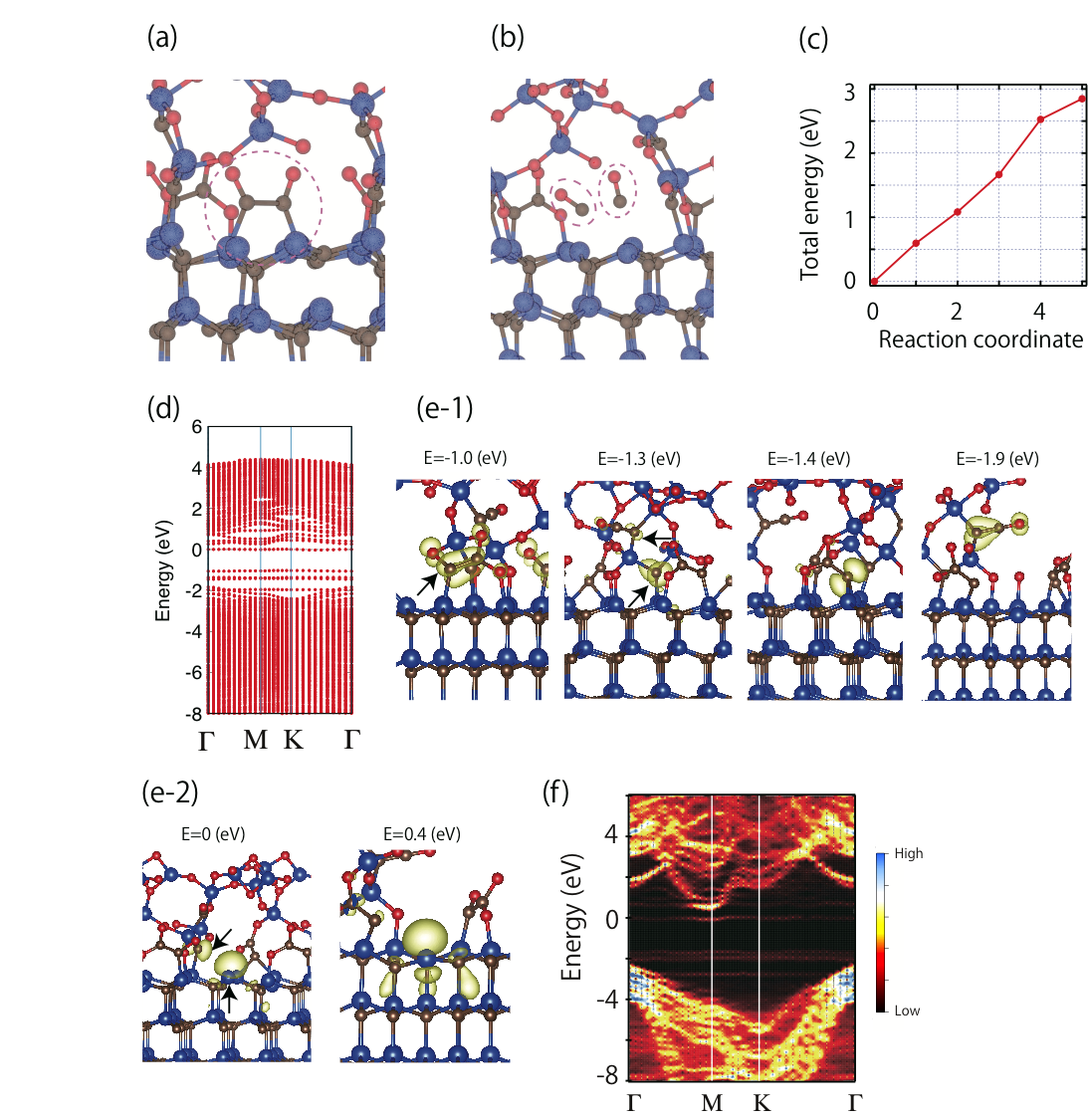}
\caption{(Color online). CO dissociation process (a) and (b) and its dissociation energy profile in (c). (d) and (f), Calculated band structures with PBE and HSE, respectively. The wave functions of the gap states are shown in Fig.~(e-1) and (e-2). The iso-value surface at 10\% of maximum values is plotted.
}
\label{Fig17}
\end{figure}

To investigate the CO$_{\rm x}$ dissociation energy from the carbon nano clusters, we have continued CPMD simulations with the temperature of 2000 [K], which is higher than the experimental temperature of thermal oxidation, in order to enhance the dissociation rate of CO$_{\rm x}$ molecules. (See the appendix \cite{CPMD_result} for the MD result). Then, during this CPMD simulation time of 30 psec., we did observe a CO dissociation once from the ethylene-like structure, Fig.~\ref{Fig15}(a). Fig.~\ref{Fig17} shows the results; In Fig.~\ref{Fig17}(a), we show the ethylene-like structure just before the CO dissociation, and Fig.~\ref{Fig17}(b) presents the structure after the CO dissociation. Now we identified a pathway of the carbon annihilation from the carbon nano clusters. Then, we took the structures before and after the CO dissociation, and performed NEB calculations after the structural optimization to evaluate the dissociation barrier (See Fig.~\ref{Fig17} (c)). The calculated energy barrier is 2.8 eV. This energy barrier corresponds to the experimental value of $\sim 3$ eV \cite{oxidation_rate1,oxidation_rate2,oxidation_rate3, oxidation_rate4}. Another important point is that the total energy of the final structure is 2.8 eV higher than the initial structure; We didn't observe the energy gain of this process, due to the appearance of dangling bonds at the Si sites of the final structure. Therefore, we passivated the dangling bonds by two (newly added) O atoms at the interface. Then, the total energy gains 3.16 eV than that of the initial structure plus O$_2$ molecule. This result shows that the termination of Si dangling bonds in interface are necessary to eject the C atoms outside and move the oxidation forwards. The calculated electronic band structure after the CO dissociation is shown in Fig.~\ref{Fig17}(d). Compared with the results of Fig.~\ref{Fig16}, the difference is the electronic states near the CBM. After the CO dissociation, the electronic state (by the PBE) derived from the ethylene-like structure Fig.~\ref{Fig15}(a) disappeared. Instead, a new electronic state appears near the CBM derived from the Si dangling bonds as shown in Fig.~\ref{Fig17}(e-2). We show the calculated band structure with the HSE in Fig.~\ref{Fig17}(f).


\section{Considerations of efficiencies of hydrogen and nitrogen treatments}\label{passiv}
We have found many possible (local minimum) defects generated during the oxidation on C- and Si-surface. Here we will discuss the reactivity of a hydrogen (H$_2$) and nitrogen (NO) molecule with the defects. 

\begin{table}[htb]
  \begin{tabular}{|c||c|c|} \hline 
  ~~~~~~~~~~~~~~~&~~~~~~~~~~H$_2$~~~~~~~~~~&~~~~~~~~~~NO~~~~~~~~~~\\[2pt] \hline \hline 
O$^{\rm surf}$&$\times$&$\times$ \\ [2pt] \hline 
Y-lid&$\times$&$\times$ \\[2pt] \hline 
Si-Si&$\bigcirc$&$\bigcirc$\\[2pt] \hline 
C$_{\rm dangling}$&$\bigcirc$&$\times$\\[2pt] \hline 
O$^{\rm inter}$&$\times$&$\times$\\[2pt] \hline 
  \end{tabular}
\caption{Efficiency of hydrogen and nitrogen treatments for the defects generated during the oxidation on C-face.}
\label{treatment_C-face}
\end{table}

First, we picked up 5 defects generated during the C-face oxidation listed in Table.~\ref{treatment_C-face}, and added an H$_2$ or NO molecule near the defects and performed structural optimization to explore whether the added species react with the defects and change the electronics structure of the defects. 
Each defect is labeled as follows; i.e., The label of "O$^{\rm surf}$" represents the defect of Fig.~\ref{Fig1}(a), the label of "Si-Si" the Fig.~\ref{Fig4} (b) structure, "C$_{\rm dangling}$" is the Fig.~\ref{Fig5} (c) structure, and "O$^{\rm inter}$" is the Fig.~\ref{Fig6_pre}(a)(b) structure. In the table, the mark "$\times$" denotes that the added species didn't react with the defect.

\begin{figure}
\includegraphics[width=0.6\linewidth]{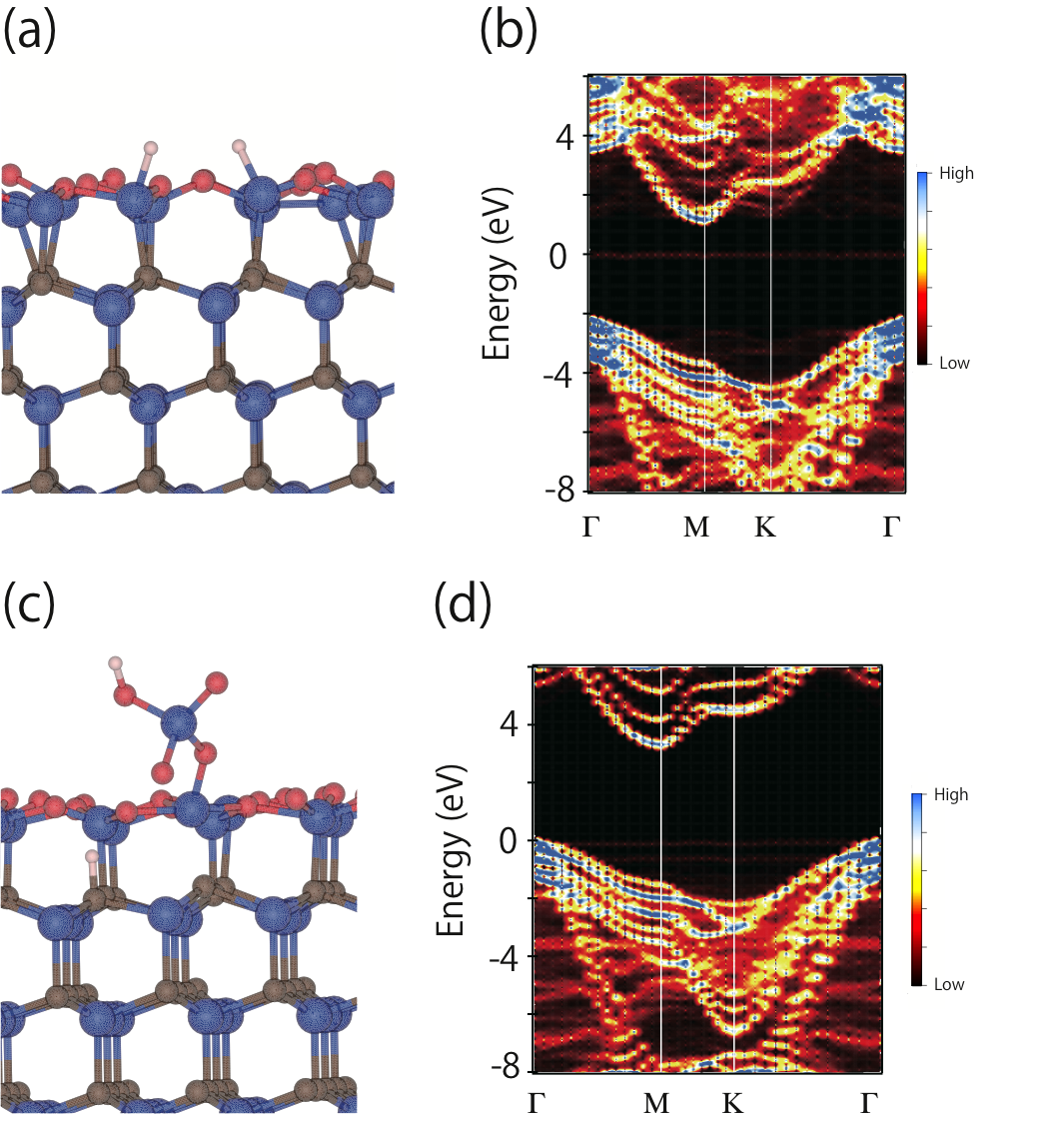}
\caption{(Color online). 
Reactivity of H$_2$ treatment with the defect structures shown in Fig.~\ref{Fig4} and Fig.~\ref{Fig5}. Structures are shown in (a) and (c). (b) and (d), Effects of H$_2$ treatment on the band structure calculated by HSE for the defects. }
\label{Fig18}
\end{figure}

We have found that for the most defects, the H$_2$ and NO treatments don't work except for "Si-Si" and "C$_{\rm dangling}$" defects. As clearly seen in the table, added H$_2$ molecule reacts with the defect of "C$_{\rm dangling}$". As we have seen, "C$_{\rm dangling}$" is thought to be an inevitable defect structure during the oxidation of C-face. We show the results of H$_2$ treatment for "Si-Si" and "C$_{\rm dangling}$" structures in Fig.~\ref{Fig18}. Fig.~\ref{Fig18} (a) shows the structure after the H$_2$ passivation for the "Si-Si" defect, and its band structure is shown in Fig.~\ref{Fig18}(b). We have found that the H$_2$ molecule terminates the Si-Si bond. Compared with Fig.~\ref{Fig4_band}, the H$_2$ treatment apparently decreases the defect level in the energy gap region. Fig.~\ref{Fig18}(c) and (d) show the results of H$_2$ passivation for the "C$_{\rm dangling}$" defect. The H$_2$ terminates the dangling bond at the subsurface C site, and the terminal O atom. The band structure is shown in Fig.~\ref{Fig18}(d). It is clearly seen that the defect levels disappear in the gap region. 


We have also investigated the reactivity of the 3 defects generated during the oxidation on Si-face with H$_2$ and NO molecules as listed in Table.~\ref{treatment_Si-face}, in which the defect in Fig.~\ref{Fig9} is called C$_2$, that in Fig.~\ref{Fig10} is C$_2$\&C$_3$, and Fig.~\ref{Fig14} "C clusters". We couldn't observe the chemical reaction of H$_2$ nor NO molecules with carbon clusters on the Si-face. 

\begin{table}[htb]
  \begin{tabular}{|c||c|c|} \hline 
~~~~~~~~~~~~~~~&~~~~~~~~~~H$_2$~~~~~~~~~~&~~~~~~~~~~NO~~~~~~~~~~\\[2pt] \hline \hline 
C$_2$&$\times$&$\times$ \\ [2pt] \hline 
C$_2$\&C$_3$&$\times$&$\times$ \\[2pt] \hline 
C clusters&$\times$&$\times$\\[2pt] \hline
  \end{tabular}
\caption{The reactivity of hydrogen and nitrogen treatments with the defects generated during the oxidation on Si-face.}
\label{treatment_Si-face}
\end{table}

\section{Conclusions}\label{sum}
We have performed electronic state calculations to clarify the initial oxidation process of 4H-SiC substrates. We have investigated how each Si and C atomic layer is oxidized on C- and Si-face, and explore most feasible reaction pathways, their energy barrier, and possible defects generated during the oxidation. We have found that carbon annihilation process is quite different between on Si- and on C-face, and this difference causes different defects in interface; In C-face case, (1), C atoms are dissociated directly from the substrate as CO molecules with 0.7 eV of dissociation barrier. (2), after CO dissociation, 3-fold coordinated oxygen atoms (called Y-lid) are observed at the interface. (3), high density of C-dangling bonds can remain at the interface. In Si-face case, (1), C atoms form carbon nano clusters (composed of a few atoms) in interface to reduce the number of dangling bonds. Moreover, we have found that the carbon nano clusters are composed of not only single but also double chemical bonds. (2), It is observed that CO molecules are dissociated from carbon nano clusters in MD simulations with 2.8 eV of dissociation energy barrier. Furthermore, we investigated whether H$_2$ and NO molecules react with the defects generated during the oxidation, which is an experimentally known technique to remedy the interface properties. Then, we have found that H$_2$ treatment efficiently reduces the C dangling bonds on C-face.

\section{Acknowledgments}
This work was supported by JSPS Grant-in-Aid for Young Scientists (B) Grant Number 16K18075.
Computations were performed mainly at the Supercomputer Center at the Institute for Solid State Physics, The University of Tokyo, The Research Center for Computational Science, National Institutes of Natural Sciences, and the Center for Computational Science, University of Tsukuba.

\end{document}